\documentclass[12pt, a4paper]{article}
\usepackage[dvipdfmx]{graphicx}
\graphicspath{{figs/}}

\usepackage{amssymb}
\usepackage{amsmath}
\usepackage{bm}
\usepackage{color}
\usepackage{slashed}
\usepackage{subfigure}
\usepackage{epstopdf}            
\usepackage{epsfig}
\usepackage[utf8]{inputenc}  
\newcommand{\beq}{\begin{equation}}
\newcommand{\eeq}{\end{equation}}

\usepackage[colorlinks=true, linkcolor=black, citecolor=black,
urlcolor=blue]{hyperref}

\begin{document}

\begin{titlepage}
\begin{center}

\begin{flushright}
MSUHEP-20-001\\
\end{flushright}
\vspace{1cm}
{\LARGE Detecting a $\mu\tau$-philic $Z^\prime$ boson\\ via photon initiated processes at the LHC}

\vskip 2cm
Syuhei Iguro$^1$, Kirtimaan A. Mohan$^2$ and C.-P. Yuan$^2$ 
\vskip 0.5cm

{\it $^1$
Department of Physics, Nagoya University, Nagoya 464-8602, Japan}\\[3pt]
{\it $^2$
Department of Physics and Astronomy,
Michigan State University, East Lansing Michigan, 48823, USA}\\[3pt]
\vskip 1.5cm

\begin{abstract}
Discrepancy between the measured value and the
Standard Model prediction of the muon anomalous magnetic moment is a possible hint for new physics.
A $Z^\prime$ particle with $\mu\tau$ flavor violating couplings can give a large contribution to the muon anomalous magnetic moment due to the $\tau$ mass enhancement at the one-loop level, and is known to explain the above discrepancy.
In this paper, we study the potential of the Large Hadron Collider (LHC) for detecting such a $Z^\prime$ boson via  the $p p \to\mu^-\mu^-\tau^+\tau^+ $ process.
Earlier studies in the literature only considered the production channel with quark initial states ($p p \to q \bar q \to\mu^-\mu^-\tau^+\tau^+ $). Here, we show that the photon initiated process, $p p \to \gamma \gamma \to \mu^-\mu^-\tau^+\tau^+ $, is in fact the dominant production mode, for a heavy $Z^\prime$ boson of mass greater than a few hundred GeV.
The potential of the high luminosity (HL) LHC is also considered. 
\end{abstract}
\end{center}
\end{titlepage}

\section{Introduction}
The long standing discrepancy in the measured value of the anomalous magnetic moment of the muon might be a hint for new  physics beyond Standard Model.
The current discrepancy is given as \cite{Blum:2018mom}\footnote{See also \cite{Hagiwara:2006jt,Jegerlehner:2009ry,Davier:2010nc,Hagiwara:2011af}. },
\begin{align}
\delta a_\mu=(2.74\pm0.73) \times10^{-9}.
\end{align}
 The muon $g-2$ experiment running at Fermilab (FNAL)~\cite{Grange:2015fou} and another planned experiment at J-PARC~\cite{Mibe:2011zz} will help to reduce the uncertainty in these measurements by a factor of four and will help to establish if $\delta a_\mu$ can indeed be taken as evidence for new physics.
A large number of new physics models have been proposed in literature to explain this discrepancy.~\footnote{ See, for example, Ref.~\cite{Lindner:2016bgg} and references therein.}
The introduction of a new particle with lepton flavor violating $\mu\tau$ couplings is known to be able to explain the observed value of $\delta a_\mu$. Examples of such models are the axion like particle model~\cite{Bauer:2019gfk,Cornella:2019uxs}, general two Higgs doublet model~\cite{Assamagan:2002kf,Davidson:2010xv,Omura:2015nja,Omura:2015xcg,Iguro:2018qzf,Abe:2019bkf,Iguro:2019sly,Wang:2019ngf} and $Z^\prime$ models~\cite{Baek:2001kca,Baek:2008nz,Heeck:2016xkh,Altmannshofer:2016brv}.

In this paper, we consider a vector boson ($Z^\prime$) which has  sizable $\mu\tau$ flavor violating couplings.
Large $\mu\tau$ flavor violating couplings are required in order to explain the muon anomaly. At the same time, flavor diagonal couplings are constrained to be small because of  stringent (low-energy) flavor physics constraints, such as $\tau\to \mu\gamma$, $\tau\to 3\mu $, and $\tau\to \mu e e$~\cite{Heeck:2016xkh}.
If flavor diagonal couplings are strongly constrained, it will be quite challenging to detect  a lepton flavor violating decay of $\tau$ via the exchange of such a $Z^\prime$ boson. 
In that case, a high-energy collider could play an important role for testing such kind of models.
Earlier studies in literature only considered the $p p \to q \bar q \to\mu^-\mu^-\tau^+\tau^+ $ production channel~\cite{Altmannshofer:2016brv}.\footnote{The charge conjugated final state $\mu^+\mu^+\tau^-\tau^-$ is also included. It is just denoted as $\mu^-\mu^-\tau^+\tau^+$ for simplicity. } 
Here, we show that the photon initiated process, $p p \to \gamma \gamma \to \mu^-\mu^-\tau^+\tau^+ $, can play an important role in the detection of a heavy $Z^\prime$ boson of mass greater than a few hundred GeVs. 
 
This paper is organized as follows. In Sec.~\ref{sec:2}, we introduce the simplified model to explain the muon anomaly.
In Sec.~\ref{sec:3}, we investigate the signal rate at the LHC and show the potential of the high luminosity HL-LHC for constraining such a $Z^\prime$ boson with an integrated luminosity of 1 or 3 ab$^{-1}$ \cite{CMS:2013xfa}.
Finally, we present our conclusions in Sec.~\ref{sec:4}.
\section{A $\mu\tau$ flavor violating $Z^\prime$ and the muon $g-2$ anomaly}
\label{sec:2}

\begin{figure}[ht]
  \begin{center}
    \includegraphics[width=5.5cm]{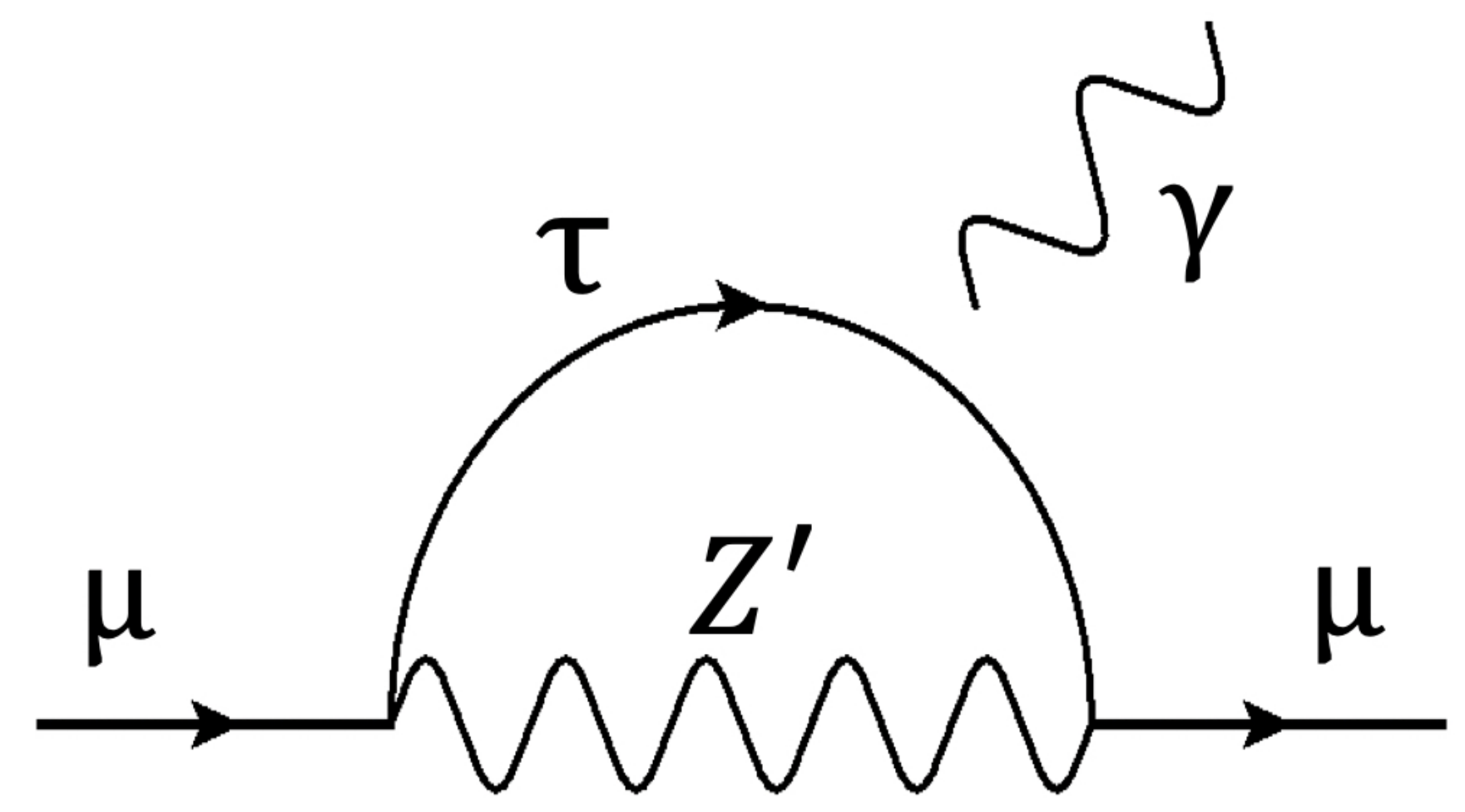}
            \caption{Relevant Feynman diagram that contributes to $\delta a_\mu$. The internal fermion is a $\tau$ lepton and mass insertions give rise to the chirality enhancement. }
    \label{diagramg-2}
  \end{center}
\end{figure}

In this section we introduce our interaction Lagrangian for a single leptophillic vector boson ($Z^\prime$) with flavor violating couplings.
Following Ref.~\cite{Altmannshofer:2016brv}, the couplings of the $Z^\prime$ boson to $\mu$ and $\tau$ leptons in the mass basis can be written down as follows:
\begin{align}
\label{Zprime}
 {\cal L_{Z^\prime}}=  \left[g^\prime_L( \bar{\mu}_{L}\gamma_{\mu} \tau_{L}+ \bar{\nu_\mu}_{L}\gamma_{\mu} \nu_{\tau L})+  g^\prime_R \bar{\mu}_{R}\gamma_{\mu} \tau_{R}\right]Z^\prime+{\rm h.c.}.
\end{align}
The purely off-diagonal terms give new contributions to the muon $g -2$. The corresponding feynman diagram for the process is shown in Fig.~\ref{diagramg-2} and its contribution to the 
anomalous magnetic moment of the muon when  $m_{Z^\prime}\gg m_\tau$, is approximately given as
\begin{align}
\label{g2_Zp}
\delta a_{\mu}^{Z^\prime}&\sim\frac{m_\mu^2}{12\pi^2 m_{Z^\prime}^2}\left[\frac{3m_\tau}{m_\mu} {\rm{Re}}\left[g^\prime_L  g^{\prime }_R\right]-|g^\prime_L|^2-|g^\prime_R|^2\right].
\end{align}
Here we will focus on $Z^\prime$ masses of the order of $100$~GeV or more and hence will only concern ourselves with the  case ($m_{Z^\prime}\gg m_\tau$).
The first term of Eq. (\ref{g2_Zp}) is enhanced by the ratio $m_\tau/m_\mu\sim 17$, which is referred to as the chirality enhancement factor. 
In order to explain the anomaly, one requires Re$[g^\prime_L g^{\prime }_R ]\ge 0$. On the other hand, a purely left or right handed coupling would lead to negative contributions to the muon $g-2$ and result in further deviation from the experimental value. 
Hence, in order to explain the anomaly we assume that both left and right handed couplings are non zero. Further, for simplicity, we assume that both $g^\prime_L$ and $g^{\prime }_R$ are real and positive.  We introduce the variable $R = g_L^\prime/ g_R^\prime$ as an independent parameter for the simplified model. The two other independent parameters being the mass of the $Z^\prime$ boson ($m_{Z^\prime}$) and the right handed coupling $g^\prime_R$.

\begin{figure}[ht]
	\begin{center}
		\includegraphics[width=5.7cm]{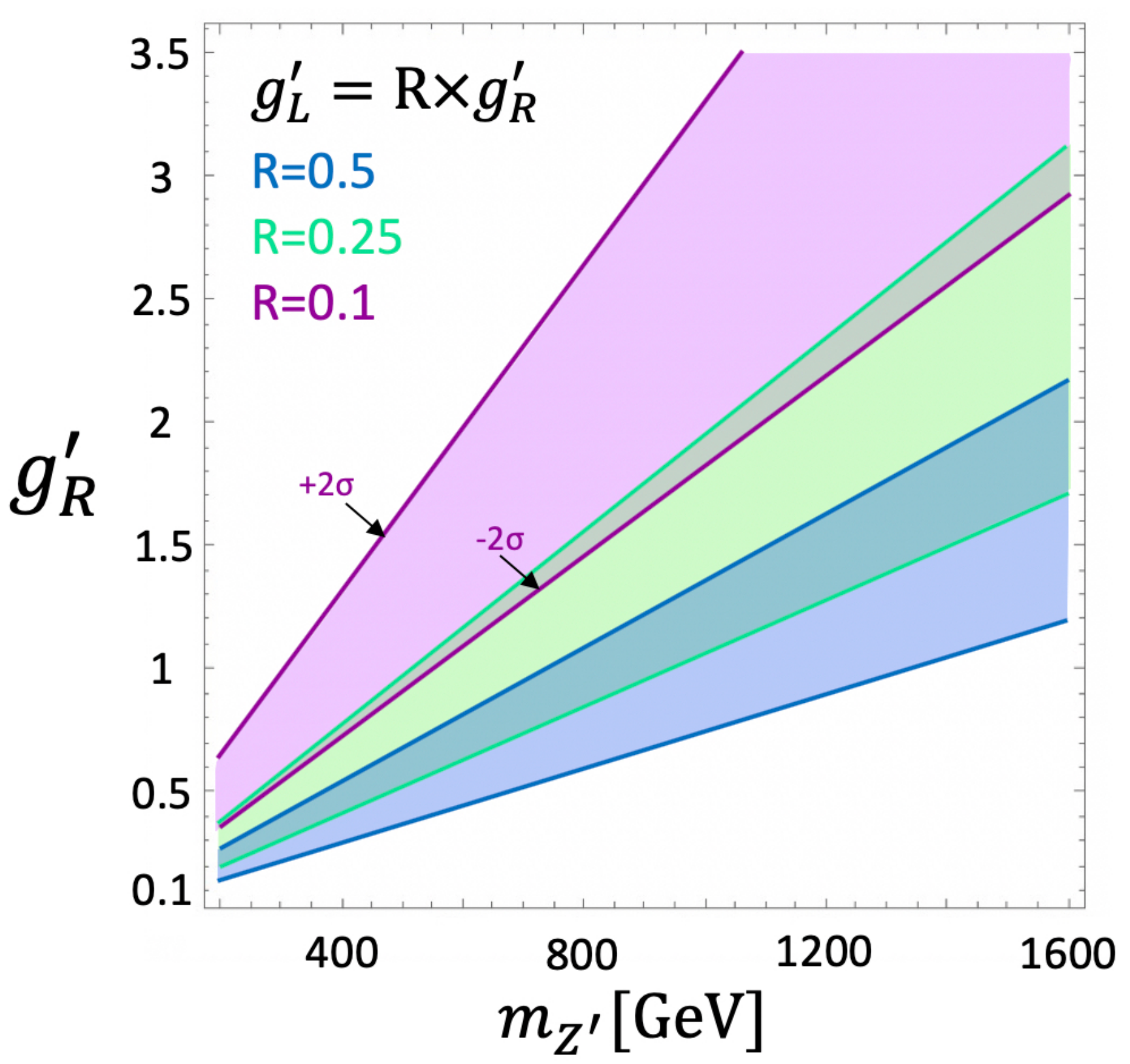}
		\caption{Parameter space of the simplified model that explain $\delta a_\mu$ within 2$\sigma$.
			The horizontal axis is for a mass of the $Z^\prime$ boson and vertical axis is for the coupling $g_R^\prime$.
			The upper and lower lines on each band indicate values of the model  parameters that accommodate the anomaly with +2$\sigma$ and -2$\sigma$ uncertainties, respectively. 
			The different colors indicate different values of the $R$.
			The blue, light green and purple bands correspond to $R=0.5$, $R=0.25$ and $R=0.1$, respectively.   }
		\label{paramZp}
	\end{center}
\end{figure}
In Fig.~\ref{paramZp}, we show a plot indicating regions of  parameter space of this simplified model that can explain the muon $g-2$ anomaly. We choose three different values of the ratio $R=\left\{0.1,0.25,0.5\right\}$. The shaded regions indicate values of the parameter space that match the  measured value of $\delta a_\mu$ to within $\pm~2\sigma$ uncertainty.
We see that $\mathcal{O}(1)$ couplings are needed in order to explain the anomaly.
Additionally, since the contribution to the anomaly is proportional to the product of left and right handed couplings, a larger $g_R^\prime$ is needed for a smaller R value. 
Finally, we ensure that the model remains perturbative by requiring that $g_R^\prime\le\sqrt{4\pi}$.
Since the couplings are constrained to be significantly large, there is a very definite possibility of LHC searches being sensitive to relevant regions of model parameter space. It is this sensitivity that we explore in detail in the following section.

\section{Collider Signal at the LHC}
\label{sec:3}
\begin{figure}[ht]
  \begin{center}
   \includegraphics[width=5.2cm]{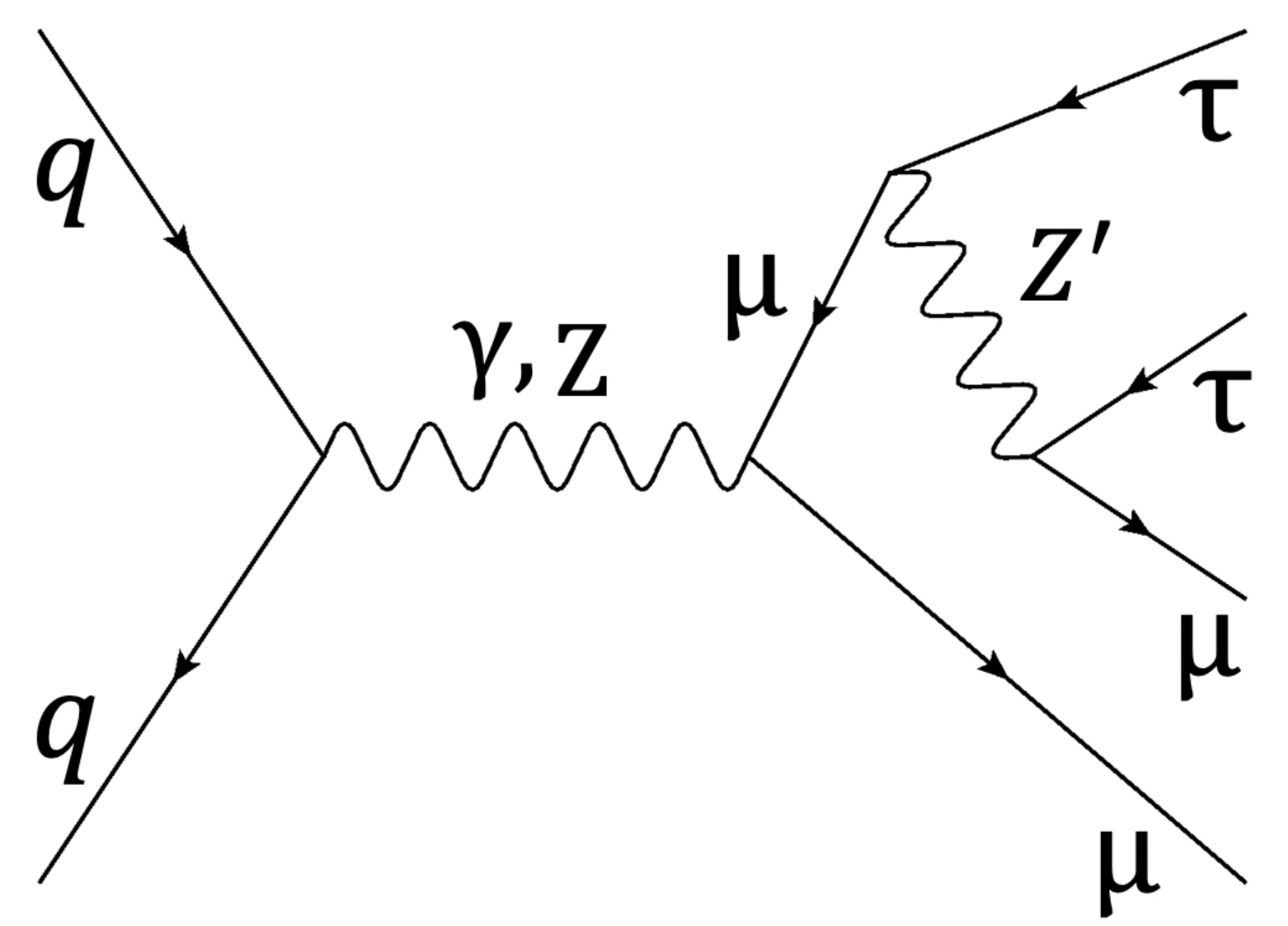}
      \includegraphics[width=4.8cm]{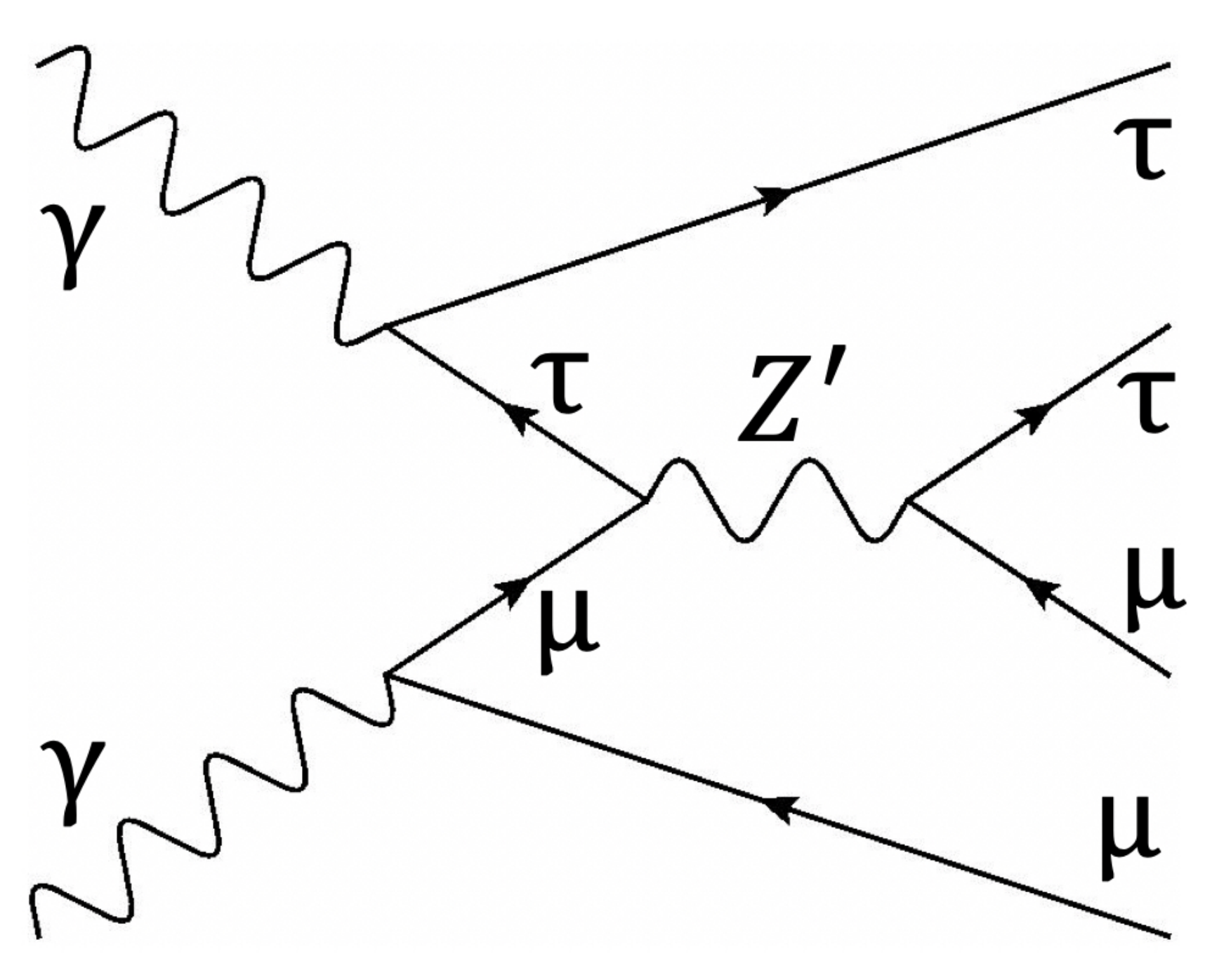}
         \includegraphics[width=4.1cm]{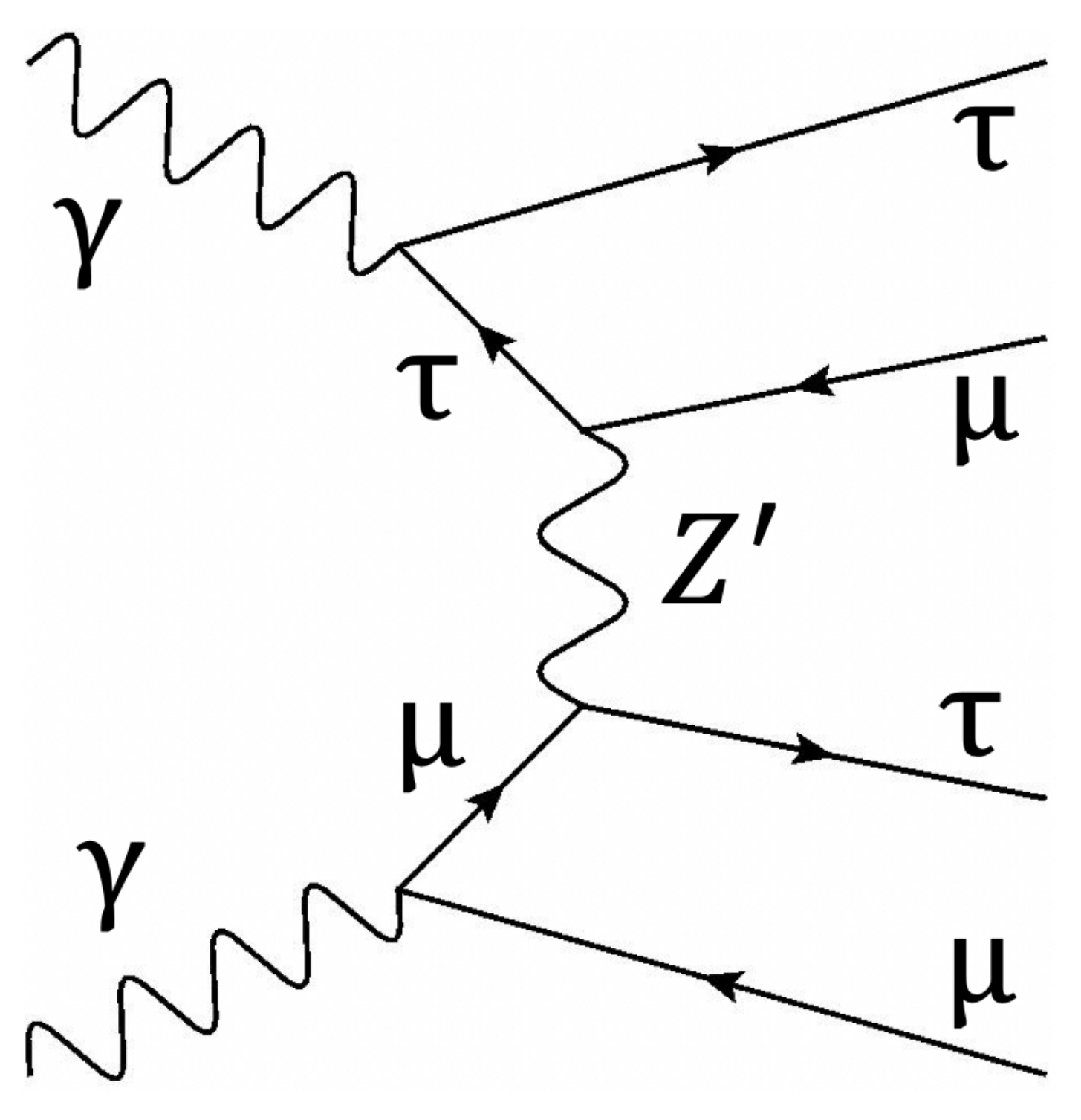}
          \caption{Representative feynman diagrams for the $\mu^-\mu^-\tau^+\tau^+$ signal at the LHC. In addition to the diagrams shown here there are also those obtained by exchanging $\mu$ and $\tau$, which are included in our numerical calculation. }
   \label{dia}
  \end{center}
\end{figure}

\begin{figure}[ht]
  \begin{center}
   \includegraphics[width=8cm]{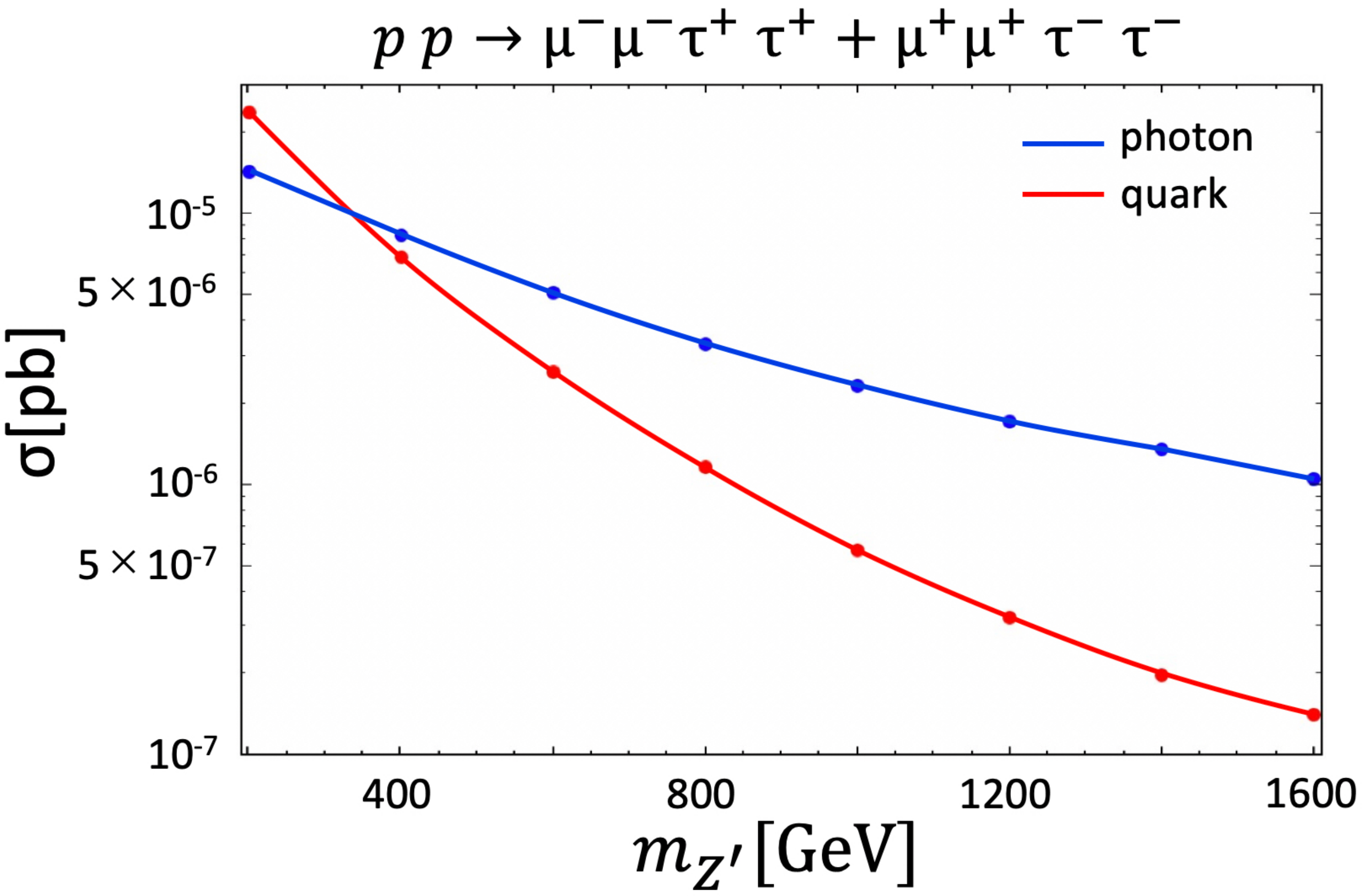}
               \caption{The total fiducial cross sections of quark-antiquark (red) and photon-photon (blue)  initiated processes of the $\mu^-\mu^-\tau^+\tau^++\mu^+\mu^+\tau^-\tau^-$ final state. The horizontal axis is the mass of the $Z^\prime$ boson in GeV and the vertical axis is the cross section in pb. Here, we have fixed the ratio of coupling to the $Z^\prime$ mass as $g_L^\prime/m_{Z^\prime}=0.3/400$ GeV$^{-1}$, with $R=1$. }
   \label{compari}
  \end{center}
\end{figure}

At a high energy proton-proton collider like the LHC, the flavor violating $Z^\prime$ boson, in the considered simplified model, can only be produced in association with a muon and tau. This results in a unique signature, namely a pair of same sign dileptons in the final state, i.e. $\mu^-\mu^-\tau^+\tau^+$ or equivalently the charge conjugate combination $\mu^+\mu^+\tau^-\tau^-$. 
In Fig.~\ref{dia}, we show the relevant feynman diagrams for the production of the $Z^\prime$ boson, which subsequently decays into $\mu$ and $\tau$ leptons. In this work, in addition to the  quark initiated process, $p p \to q\bar q \to\mu^-\mu^-\tau^+\tau^+ $, we also take into account the photon initiated processes, $p p \to \gamma \gamma \to\mu^-\mu^-\tau^+\tau^+ $, that has not been considered before. In fact, we will find that for most regions of parameter space, it is the photon initiated process that dominates the production of the $Z^\prime$ boson.
This fact can be seen in Fig.~\ref{compari} which shows the total fiducial cross sections of both 
the quark-antiquark (red) and photon-photon initiated (blue) processes in the $\mu^-\mu^-\tau^+\tau^+$ final state. 
In Fig. 4, we have applied the following cuts to the final state leptons: 
\begin{enumerate}
	\item Transverse momentum of all leptons $p_{T_l} > 10 ~\rm{GeV}$.
	\item Pseudo-rapidity of all leptons $|\eta| <2.5 $.
	\item Separation between all leptons $\Delta R(l,l) > 0.4$.
\end{enumerate}
Here, we set $R =1$ and the value of the coupling $g_R^\prime = m_{Z^\prime} /\left(400[{\rm GeV}]\right)$, as a demonstration.  
The calculation was performed with the help of {\tt MadGraph\_amc@NLO}~\cite{Alwall:2014hca},
 while the necessary model file was produced using {\tt Feynrules}~\cite{Alloul:2013bka}.
We use {\tt NNPDF\_31\_nnlo\_luxqed}~\cite{Bertone:2017bme} parton distribution function (PDF) to determine both the quark and photon PDFs  and set both the renormalization and factorization scales to be equal to the mass of the $Z^\prime$ boson.

As can be seen from the figure, the dominant contribution to the signal cross section for high masses ($m_{Z^\prime}\ge 400$ GeV) comes from the photon initiated diagrams.\footnote{Our cross sections for the quark initiated process are smaller than those in Ref.~\cite{Altmannshofer:2016brv}. }
This is largely due to the behavior of the photon PDF at large values of the parton  momentum  fraction $(x)$~\cite{Bertone:2017bme}. When taking the photon as a parton inside the proton, the photon PDF becomes larger than the quark PDF as the parton momentum fraction $x$ approaches a value of  1~\cite{Martin:2004dh,Ball:2012cx,Ball:2014uwa,Schmidt:2015zda,Bertone:2016ume,Manohar:2016nzj,Manohar:2017eqh}.

Recall that in order to explain the muon $g-2$ anomaly, large values of the couplings $g_L^\prime$ and $g_R^\prime$ are required. It is therefore important to check carefully that the width of the resonance is not too large in order to perform the cross section calculation reliably.
Neglecting the masses of $\mu$ and $\tau$ leptons, the decay width $\Gamma$ of the $Z^\prime$ boson can be written as follows:
\begin{align}
\frac{\Gamma }{m_Z^\prime}=\frac{\left(g_R^\prime\right)^2\left( 1 + R^2\right)}{12\pi}\ . 
\end{align}
When the magnitude of the right handed coupling is near the perturbative limit ($g^\prime_R \simeq\sqrt{4\pi}$), the ratio is given as, $\Gamma/ m_{Z^{\prime}}\simeq (1+ R^2) /3$. 
From Fig.~\ref{paramZp} we see that the values of $R$ are constrained to be small ($\sim 0.1$) when $g_R^\prime$ is large.  For large values of this ratio ($\Gamma/m_{Z^\prime}> 0.3$) the accuracy of the Breit-Wigner approximation with a fixed width drops and one needs to take into account a running width that depends on the energy of the $Z^\prime$.\footnote{See, for example, Ref. \cite{Chivukula:2017lyk} and references therein.}
However, for the region of parameter space of interest, the largest values that we need concern ourselves with is $\Gamma/m_{Z^\prime} \sim 0.3$ and so we use the Breit-Wigner approximation with constant $\Gamma$ for simplicity.
\begin{figure}[t]
  \begin{center}
    \includegraphics[width=10cm]{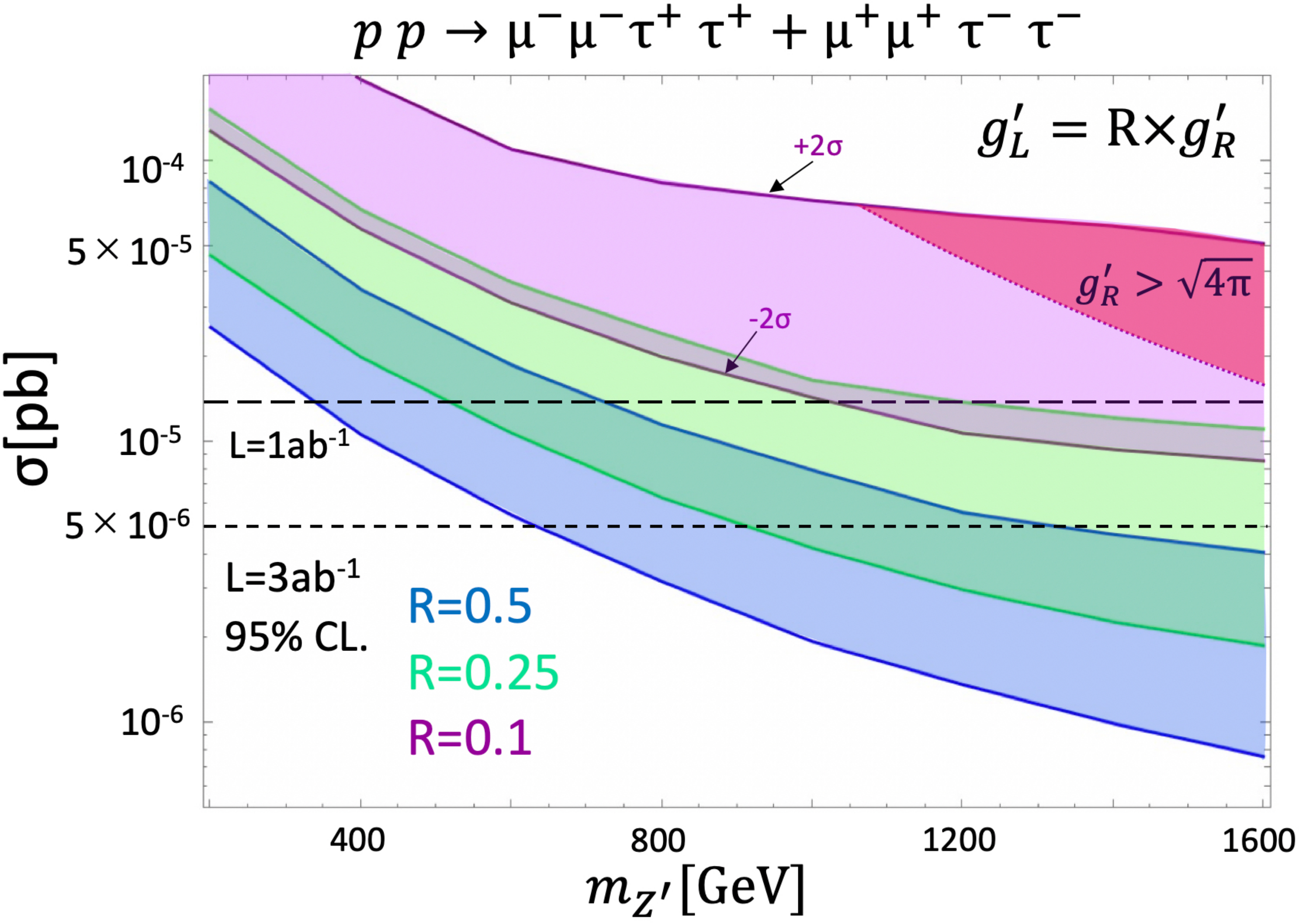}
            \caption{The figure shows future sensitivity for the $Z^\prime$ masses in the range of $400$~GeV to $1600$~GeV.
            The horizontal and vertical axes correspond to the $Z^\prime$ mass, in GeV, and the cross section, in pb,  of the $\mu^-\mu^-\tau^+\tau^++\mu^+\mu^+\tau^-\tau^-$ final state, including photon initiated processes. Each color band indicates the prediction of the cross section: $R=0.5$ (blue), $R=0.25$ (light green), and $R=0.1$ (purple).
            The red region corresponds to the parameter space of $g_{R}^\prime\ge\sqrt{4\pi}$. 
            The upper (lower) line corresponds to the cross section for model parameters that explain the muon anomaly to within 2$\sigma$ above (below) of the central value.
            The black long (short) dashed line is the sensitivity with 1 (3) ab$^{-1}$ of the data at the 95$\%$ CL assuming Poissonian statistics. 
            }
    \label{ResultZp}
  \end{center}
\end{figure}

An interesting feature about the $\mu^-\mu^-\tau^+\tau^++\mu^+\mu^+\tau^-\tau^-$ signal is that the Standard Model background to this process is negligible. It is thus possible to estimate the sensitivity of the LHC to such a process simply by calculating the fiducial cross section of the process and to determine the number of events as $N=L\times\sigma\times\epsilon$, where $L$, $\sigma$, $\epsilon$ are luminosity, fiducial cross section and product of lepton tagging efficiencies respectively. 
We apply the same cuts as those introduced earlier to draw Fig.~\ref{compari}.
Since we have only a small number of signal events with no background, we must use Poisson statistics.
Exclusion at 95$\%$ confidence level (CL) occurs when 3 signal events are predicted for a choice of model parameters, while no events are observed experimentally~\cite{Tanabashi:2018oca}.
We assume that the $\tau$ tagging efficiency is $70\%$ and only consider hadronically decaying $\tau$ leptons, with a decay branching ratio of BR($\tau\to $ hadrons )=0.65.
Therefore the efficiency is given as $\epsilon\sim 0.21$ and the cross section that can be excluded at 95$\%$ CL with luminosity of 1 (3) ab$^{-1}$ is $\sigma^\prime=1.5$ $(0.5)\times10^{-5}$ pb. 

Fig.~\ref{ResultZp} shows the cross section of the signal in pb as a function of the  $Z^\prime$ mass in GeV. 
The signal cross section of the $\mu^-\mu^-\tau^+\tau^++\mu^+\mu^+\tau^-\tau^-$ final state includes both the quark-antiquark and photon-photon fusion processes. Each colored band shows predicted values of the cross section for model parameter choices that can explain the muon $g-2$ anomaly (cf. Fig.~\ref{paramZp}). As usual, different colors indicate different values of $R$;   $R=0.5$ (blue), $R=0.25$ (light green), and $R=0.1$ (purple). We also indicate in the plot in red, the region of parameter space $g_{R}^\prime\ge\sqrt{4\pi}$, which is not allowed by the perturbativity requirement.  The upper (lower) line corresponds to cross sections for the parameters that explain the $g-2$ anomaly to within $ 2 \sigma$ above (below)  the experimentally measured central value. 
The black long (short) dashed line indicates sensitivity of the 14 TeV HL-LHC with 1 (3) ab$^{-1}$ of integrated luminosity at the 95$\%$ CL, using Poisson statistics as explained previously. 
Fig.~\ref{ResultZp} shows that the HL-LHC can exclude a  $Z^\prime$ of mass up to 350 (650) GeV when $R=0.5$  using  1 (3) ab$^{-1}$ of the data.
Additionally, HL-LHC with  3 ab$^{-1}$ of the data  is sensitive to and can constrain all the $Z^\prime$ mass parameter region shown in Fig.~\ref{ResultZp} when $R=0.1$. 
As expected, the weakest constraints from the LHC corresponds to the choice $R=0.5$, for which the $g_R^\prime$ couplings are the smallest.
\begin{figure}[t]
  \begin{center}
    \includegraphics[width=5.35cm]{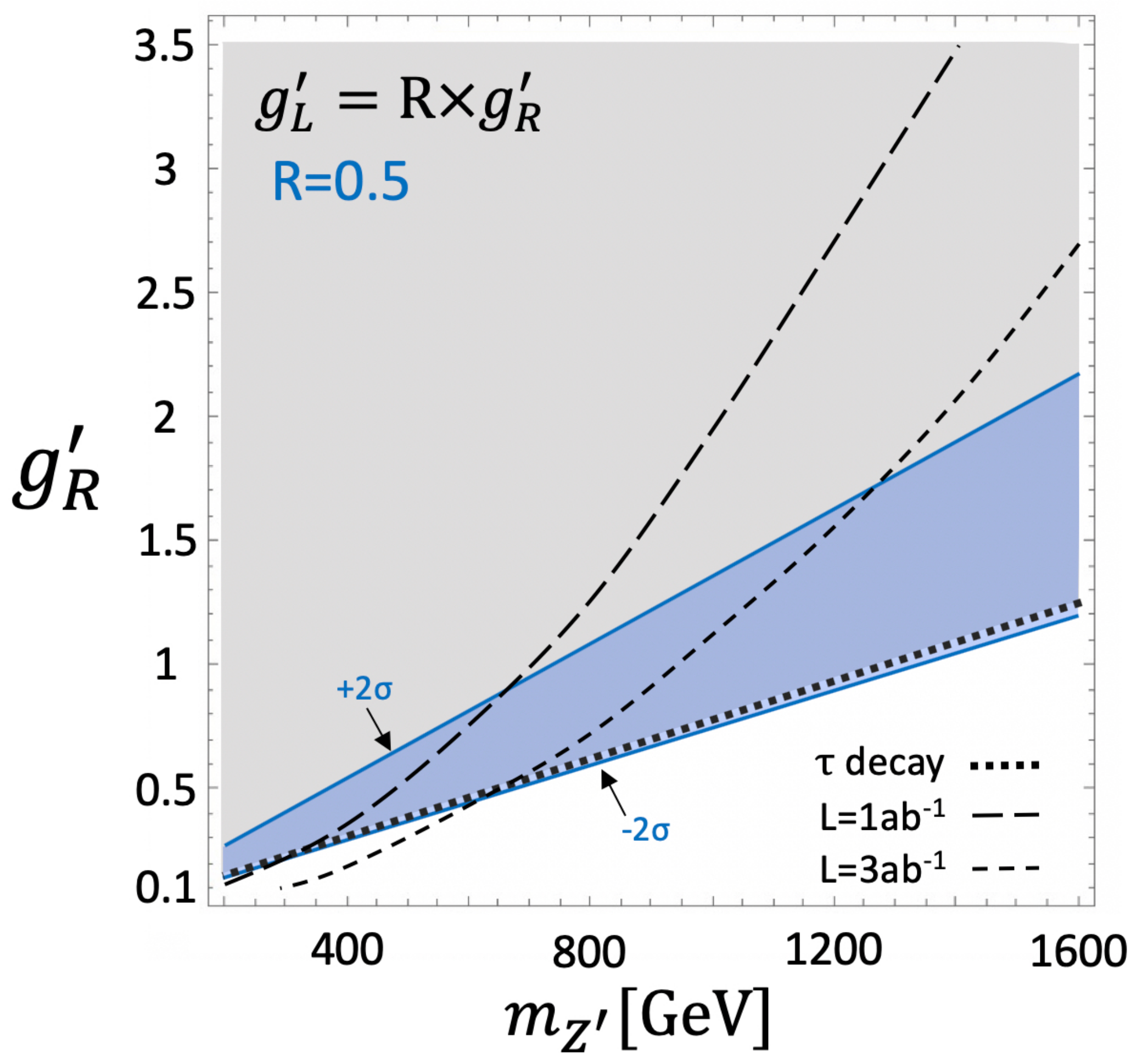}
    \includegraphics[width=5.35cm]{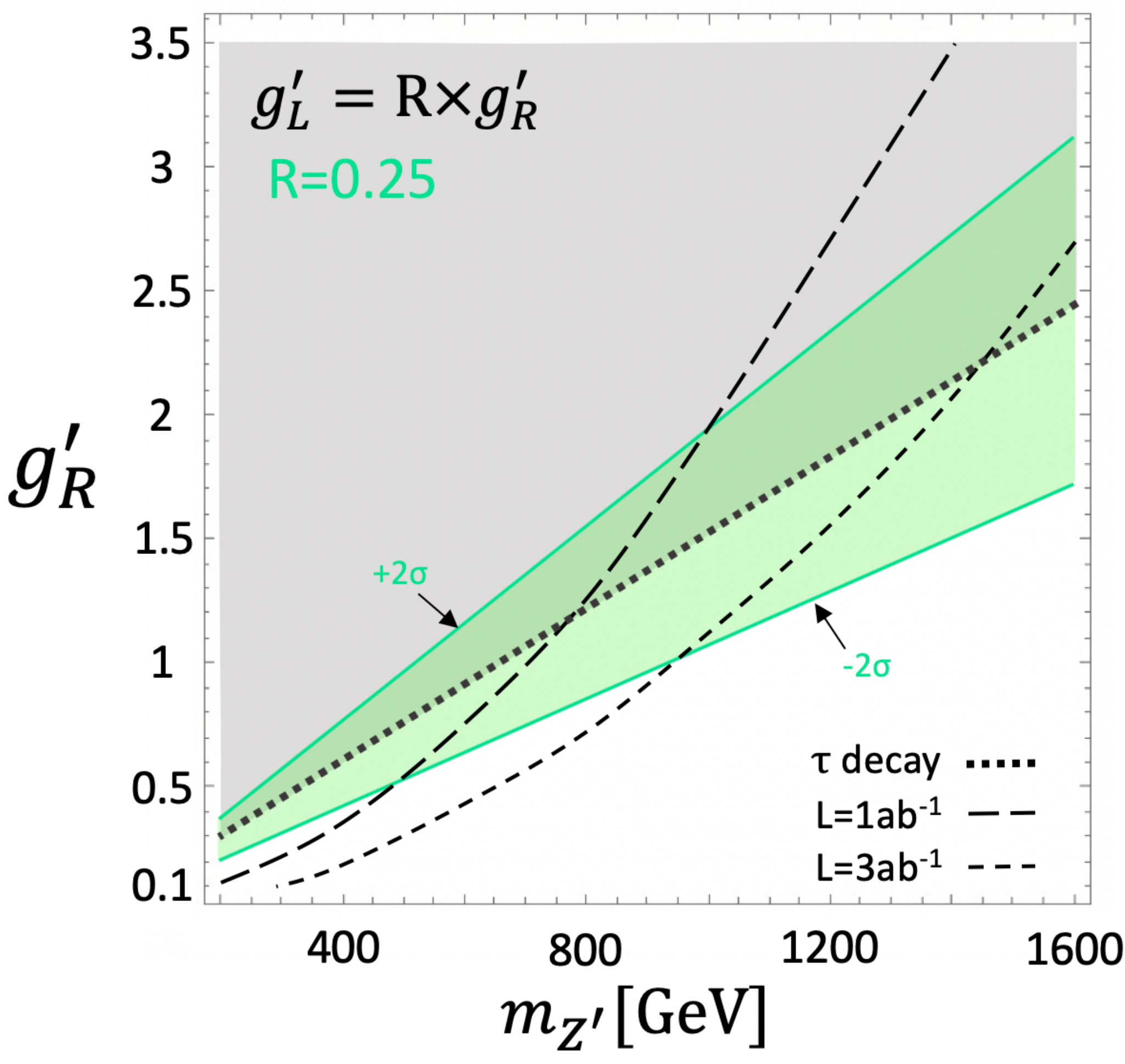}
    \includegraphics[width=5.35cm]{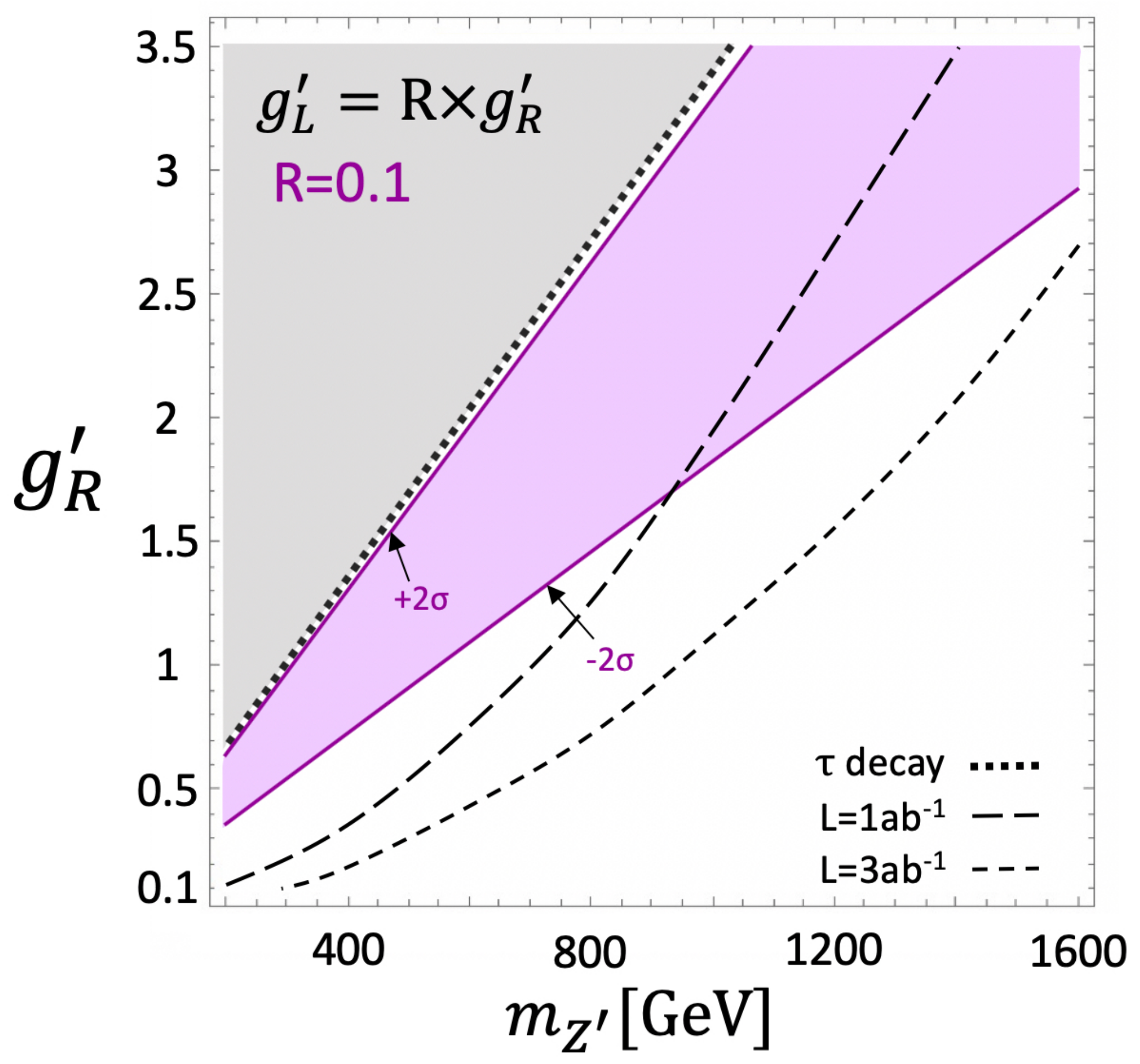}
            \caption{The sensitivity plots in a $g^\prime_R$ vs $m_{Z^\prime}$ plane are shown for $R=0.5$ (left), $R=0.25$ (center) and $R=0.1$ (right).
	The black long (short) dashed lines show the future sensitivity of the HL-LHC with 1 (3) ab$^{-1}$ of integrated luminosity at the 95$\%$ CL.
	The region above the dashed lines will be constrained by the HL-LHC data.
	We also draw dotted lines to show the constraint from BR($\tau\to\mu\nu\nu$) in each panel.
	 The region above each dotted line (the grey shaded region) is excluded.
	 The colored region bounded by two straight lines is consistent with the muon $g-2$ data within 2$\sigma$, for a given choice of the $R$ value.                        }
    \label{ResultPara}
  \end{center}
\end{figure}

In order to show all model parameters and experimental constraints simultaneously, we 
recast the potential of the HL-LHC for the detection of such a $Z^\prime$ boson 
in the mass versus coupling plane as  shown in Fig.~\ref{ResultPara}.
Further, we include constraints that arise from the three body decay of $\tau\to\mu\nu\nu$.
We use the latest bounds on BR($\tau\to\mu\nu\nu$) that can be found in Ref.~\cite{Tanabashi:2018oca}.\footnote{Expressions for the decay width of $\tau\to\mu\nu\nu$ can be found in Ref.~\cite{Altmannshofer:2016brv}.} In Fig.~\ref{ResultPara}, the 95$\%$ CL bound from $\tau$ decays rules out the region above the dotted line in each scenario (the grey shaded region).
For the $R=0.5$ scenario, nearly the entire region of parameter space, viable for explaining  the muon $g-2$ anomaly to within 2$\sigma$ (the colored region bounded by the two straight lines), has already been excluded by the BR($\tau\to\mu\nu\nu$) constraint. 
When the ratio $R$ becomes smaller, constraints imposed by the BR($\tau\to\mu\nu\nu$) data becomes weaker.
For example, for the $R=0.1$ scenario, none of the parameter space, viable for explaining  the muon $g-2$ anomaly within the 2$\sigma$ level is excluded by the BR($\tau \to \mu \nu\nu$) data.
This is because the decay rate is proportional to $|g_L^\prime|^2$. 
On the other hand, when $R$ is small, larger $g_R^\prime$ is needed to explain the muon $g-2$ anomaly.
As a result, a large signal cross section is expected at the LHC when the value of  $R$ is small.
We indicate on Fig.~\ref{ResultPara} by the black long (short) dashed lines,  constraints from the HL-LHC  with 1 (3) ${\rm ab}^{-1}$ integrated luminosity.
Regions of parameter space above these lines will be constrained at the $95\%$~CL level.
We see that the entire region of the parameter space will be ruled out by the HL-LHC data for $R=0.1$, while small but significant regions of parameter space will remain unconstrained by all the experimental data for $R=0.25$. 
Fig.~\ref{ResultPara} demonstrates the interesting complementarity between  direct searches at the LHC and the $\tau$ decay constraint obtained at low energy colliders.
While large values of $R$ are constrained by the $\tau$ decay data, small values of $R$ are more strongly constrained by the HL-LHC.
We see that even after the HL-LHC run there will be small regions of the parameter space which are consistent with both the muon $g-2$ anomaly and the BR($\tau\to\mu\nu\nu$) measurement.
For example, parameter space points with $R=0.25$, $m_{Z^\prime}>1500$~GeV and $g^\prime_R~\sim 1.5$ will remain unconstrained. 

\section{Summary and discussion}
\label{sec:4}

The anomaly in the experimental measurement of the  muon anomalous magnetic moment is a possible and exciting hint of new physics.
Here, we have introduced a simplified model with a $\mu\tau$ flavor violating $Z^\prime$ boson, that has both right and left handed couplings.
This model can explain the anomaly due to the $\tau$ mass enhancement. 
When the right-handed coupling is much larger than the left-handed one, the data from low-energy flavor physics experiments is less constraining and large parameter space is still available.
Such a region of parameter space predicts the unique high-energy collider signal of $\mu^-\mu^-\tau^+\tau^++\mu^+\mu^+\tau^-\tau^-$ production at the LHC.
We evaluated the sensitivity of the LHC for probing the parameter space of this simplified model by taking into account photon initiated processes, in addition to quark-antiquark fusion processes. We found that for most regions of parameters space of interest, the photon initiated process dominates and the quark initiated process can be an order of magnitude smaller.
We found that large regions of the model parameter space can be tested at the HL-LHC with 3 ab$^{-1}$ of integrated luminosity. 

Further, we also demonstrated the complementarity of the tau decay BR data and the LHC direct search results in constraining relevant regions of the parameter space.
While low energy experiments can provide hints of new physics, if the $Z^\prime$ boson is not too heavy, it may be possible to reconstruct its mass at the LHC.
Thus providing further complementarity between the LHC and the low energy experiments.
It will be interesting to similarly explore the sensitivity of the Belle II experiment through beamstrashlung and brehmstrahlung processes as well as the sensitivity of future hadron colliders.

\section*{Acknowledegments}
The authors thank Yuji Omura, Michihisa Takeuchi and Kazuhiro Tobe for valuable discussions and comments. 
The work of S. I. is supported by Kobayashi-Maskawa Institute for the Origin of Particles and the Universe, Toyoaki scholarship foundation and the Japan Society for the Promotion of Science (JSPS) Research Fellowships for Young Scientists, No. 19J10980.
S. I. would like to thank all the warm hospitality during his stay at Michigan State University.
This work was also supported by the U.S. National Science Foundation under Grants No. PHY-1719914.
C.-P. Yuan is also grateful for the support from
the Wu-Ki Tung endowed chair in particle physics.

\clearpage\newpage

\bibliography{refs}
\bibliographystyle{utphys}

\end{document}